\documentclass[12pt]{article}
\usepackage{amsmath,amssymb}

\usepackage{graphicx}
\usepackage{wrapfig}
\usepackage{color}

\usepackage{hyperref}
\usepackage[sort,compress]{cite}

\def\beq{\begin{eqnarray}}
\def\eeq{\end{eqnarray}}

\def\ba{\bar{a}}
\def\bb{\bar{b}}
\def\lb{\label}
\def\om{\Omega}

\newcommand{\be}{\begin{equation}}
\newcommand{\ee}{\end{equation}}
\newcommand{\bea}{\begin{eqnarray}}
\newcommand{\eea}{\end{eqnarray}}
\newcommand{\bg}{\begin{gather}}

\newcommand{\bseq}{\begin{subequations}}
\newcommand{\eseq}{\end{subequations}}

\renewcommand{\ln}{\mathop{\rm ln}\nolimits}

\def\be{\begin{eqnarray}}
\def\ee{\end{eqnarray}}
\def\lb{\label}

\begin{document}

\title{\vspace{-0.8cm}\textbf{
The fate of black hole horizons \\
in semiclassical gravity}}

\author{ \normalsize{\textbf{ Cl\'ement Berthiere$^{1}$, Debajyoti Sarkar$^{2}$ and Sergey N. Solodukhin$^{1}$ }}} 

\date{}
\maketitle
\begin{center}
\emph{$^{1}$Laboratoire de Math\'ematiques et Physique Th\'eorique CNRS-UMR
7350 }\\
  \emph{F\'ed\'eration Denis Poisson, Universit\'e Fran\c cois-Rabelais Tours, }\\
  \emph{Parc de Grandmont, 37200 Tours, France}\\

\vspace{0.4cm}
\emph{$^{2}$Albert Einstein Center for Fundamental Physics}\\
  \emph{Institute for Theoretical Physics }\\
  \emph{University of Bern, Switzerland}
\end{center}




\vspace{0.2mm}

\begin{abstract}

\noindent { 
The presence of a horizon is the principal marker for black holes as they appear in the classical theory of gravity.
In General Relativity (GR), horizons have several defining properties. First, there exists a static spherically symmetric solution to vacuum Einstein equations which possesses a 
horizon defined as a null-surface on which the time-like Killing vector becomes null. 
 Second, in GR,
a co-dimension two sphere of minimal area is necessarily a horizon. 
On a quantum level, the classical gravitational action is supplemented by the  quantum effective action obtained by integrating out 
the quantum fields propagating on a classical background. 
In this note we consider the case when the quantum fields are conformal and perform a certain non-perturbative analysis of the semiclassical equations obtained by varying the complete gravitational action. 
We show that, for these equations,
both of the above aspects do not hold. More precisely, we prove that i) a static spherically symmetric metric that would describe a horizon with a
finite Hawking temperature is, generically, {\it not} a solution; ii) a minimal $2$-sphere
is {\it not} a horizon but a tiny throat of a wormhole. 
We find certain bounds on the norm of the Killing vector at the throat and show that it is, while non-zero, an exponentially small
function of the Bekenstein-Hawking (BH) entropy of the classical black hole. We also find that the possible temperature of the semiclassical geometry is exponentially small for large black holes. These findings suggest that a black hole in the classical theory can be viewed as a certain (singular) limit of the semiclassical wormhole geometry.
We discuss the possible implications of our results.
}

\end{abstract}

\vskip 1 cm
\noindent
\rule{7.7 cm}{.5 pt}\\
\noindent 
\noindent
\noindent ~~~ {\footnotesize e-mails: clement.berthiere@lmpt.univ-tours.fr,\hspace{2mm}sarkar@itp.unibe.ch,\hspace{2mm}sergey.solodukhin@lmpt.univ-tours.fr.}

\pagebreak

\section{Introduction}

Existence of black holes is one of the most fascinating predictions of Einstein's theory of gravity.
There is accumulating astrophysical evidence that black holes, or compact objects that look pretty much
like black holes, are not rare in the Universe. The catalog \cite{Corral-Santana:2015fud} of stellar-mass black holes contains hundreds 
of candidates. Supermassive black holes are believed to be in the center of 
any galaxy including the Milky Way. Yet another evidence comes from the recent detection of a gravitational wave signal 
which originates from a coalescence of two massive black holes \cite{Abbott:2016blz}. However, the direct detection of a black hole 
event horizon remains the principal experimental challenge.

On the other hand, there have been suggestions \cite{Solodukhin:2004rv}, \cite{Damour:2007ap}, \cite{Barausse:2014tra}, \cite{Nakamura:2016gri},
\cite{Germani:2015tda}
that wormholes may mimic very closely the behavior of black holes,
including the geodesics and the characteristic quasi-normal modes, although not having the defining
property of black holes - the existence of a horizon. The latter is replaced by a tiny throat, the longitudinal size of which is such that
it may serve as a storage for the information ever fallen into the ``black hole''. For sufficiently small deviation
parameter, the wormhole geometry is very difficult, if ever possible, to distinguish experimentally from the 
black hole geometry. In particular, as was discussed in \cite{Cardoso:2016rao}, the gravitational ringdown encoded in the shape
of the recently observed gravitational wave signal is not sufficient to actually probe the horizon and, thus, distinguish the two geometries.

The goal of this note is to provide more evidence for the wormhole picture and to demonstrate that, 
on the theoretical side, the existence of black holes, as we know them in General Relativity, is
far from evident as soon as the quantum modifications of GR are taken into account.
Indeed, the quantum fields generate a certain, generally non-local, modification (see for instance \cite{Barvinsky:1985an}) of the gravitational action and
a respective modification of the gravitational equations. In a semiclassical description, in which
the gravitational field (metric) is not quantized and all other matter fields are considered to be quantum,
the modified Einstein equations define
the so-called semiclassical gravity. These modified equations, being fundamentally non-local, are extremely complicated so that
the exact solutions can be found only in some very special, symmetric, cases \cite{Solodukhin:1998ph}, \cite{Kaus:2009cg}.
Previous works on semiclassical black holes include \cite{previous}.

There are two aspects of classical horizons as they appear in GR. First of all, a horizon is simply a special surface
in a static spherically symmetric metric on which the time-like Killing vector becomes null.
The non-vanishing gradient of the norm of the Killing vector at the horizon defines the Hawking temperature.
To leading order, near the horizon, the Einstein equations are satisfied for any temperature.
The latter is fixed to be related to the mass by considering the solution globally, everywhere between the horizon and spatial
infinity.
 
The second aspect relates horizons to surfaces of minimal area. Indeed, in GR if a co-dimension two sphere $\Sigma$ is minimally embedded in 
a four-dimensional static spacetime then this sphere is a horizon (or, mathematically more rigorously, a bifurcation sphere of the event horizon).
This second aspect is less known so that we will review it below. 

Our main goal in this note is to analyze both these aspects in the framework of a semiclassical theory of gravity. To simplify the analysis
we shall consider the quantum modification of the gravitational action produced by quantum conformal fields. 
In this case the scaling properties of the quantum action are uniquely fixed by the conformal charges of the CFT.
This helps to make a rather general analysis for an arbitrary unitary CFT. Yet, our analysis is essentially local: we expand the metric
in a small vicinity of the would be horizon  and analyze the local solution to the modified gravitational equations.
Thus, we do not have access to the global behavior of the solutions. 
 However, this local analysis happens to be extremely informative as it allows us to rule out solutions with horizons,
 and, in fact, detect the drastic deviations from the classical behavior. More precisely, we have found that
 
 \medskip
 
 \noindent i) a static spherically symmetric metric with a horizon characterized by a finite (non-vanishing) 
 temperature is, generically, {\it not} a solution to the semiclassical gravitational equations;\footnote{This statement is not to be confused with the smooth horizon structure of the well-understood Hartle-Hawking state for Rindler spacetime or similar set-ups. As we comment in remarks \ref{stateremark} of section \ref{sec:remarks}, we are exclusively looking at states (such as Boulware), where the stress-tensor does diverge near the horizon and as a result modifies the geometry there.}
 
 \medskip
 
 \noindent ii) in semiclassical gravity, a $2$-sphere of minimal area embedded in a static spacetime is {\it not} a horizon.
 Rather, it is a throat of a wormhole. We find a bound on the norm of the Killing vector at the throat and show that it is an exponentially small function of the Bekenstein-Hawking entropy
 of the classical black hole.

\medskip
 
 Thus, the static solutions to the semiclassical gravity are {\it horizonless} and the classical horizons are replaced by wormholes! 
 This is as anticipated in \cite{Solodukhin:2004rv}.
 Our result ii) shows that for the astrophysical black holes, the parameter (the smallest value of the norm of the Killing vector)
 that characterizes the deviation of the wormhole geometry from that of a black hole, although non-vanishing, is extremely small.
 That is why it might be extremely difficult to detect the deviation experimentally. Below we discuss this and other implications of our findings.
 We stress that our results concern only the static configurations. Although we anticipate that they can be extended to a stationary, rotating case,
 we can not exclude that there may exist some dynamical, time dependent solutions with an evolving horizon.

\section{Two aspects of horizons in GR}
We consider a static spherically symmetric metric of the general form (we prefer to work in  the Euclidean signature)
\be
ds^2 = \om^2(z) g_{\mu\nu}dx^\mu dx^\nu=\Omega^2(z)\left(dt^2 + N^2(z)dz^2 + R^2(z)(d\theta^2+\sin^2\theta\,d\phi^2)\right)\,.
\lb{1}
\ee
The geometrical radius of a 2-sphere is $r(z)=R(z)\Omega(z)$.
Upon varying the gravitational action with respect to $\Omega(z)$, $N(z)$ and $r(z)$ one gets three equations, one of which by Bianchi identities follows from the other two. The norm of the time-like Killing vector $\xi=\partial_t$ is $\xi^2=\Omega^2(z)$. Vanishing of this norm signals the existence of a horizon.
Clearly, this is a point (or in fact a 2-sphere) where the function $\Omega(z)$ vanishes. A particular choice for the function $N(z)$ is a matter of convenience and a choice of the coordinate system.

\medskip

\noindent{\bf I. Universality near horizon.} Consider the gauge $N(z)=1$. 
 Assuming that there exists a horizon at $r=r_h$ with a finite temperature $T=1/\beta$, one finds the near horizon behavior 
\be
\Omega(z)=e^{-2\pi z/\beta}+ \dots \, , \ \ R(z)=r_h e^{2\pi z/\beta}+\dots \, ,
\lb{2}
\ee
where $\dots$ stand for subleading terms. The regularity of the metric requires the Euclidean time to be periodic with period $\beta$. Notice that in these coordinates the horizon is located at $z\rightarrow \infty$. In this regime, the optical metric $g_{\mu\nu}$  in (\ref{1})
universally approaches \cite{Sachs:2001qb} a product space of  one-dimensional circle $S_1^{\beta}$ with a $3$-dimensional hyperbolic space $H_3$ of radius $\beta/(2\pi)$.
\medskip

\noindent {\bf Horizon in classical theory (aspect A):} {\it There exists an exact solution to the classical Einstein equations (with or without cosmological constant) such that
the static spherically symmetric metric locally behaves as in (\ref{2}).} To leading order, the equations are satisfied for any $\beta$. The latter is related to the mass by studying the solution globally. 

\medskip

\noindent{\bf II. Horizon as a minimal surface.} Consider now the gauge $N(z) = 1/\Omega(z)$.  
In this case the radial coordinate $\rho=z$ measures the geodesic distance in the radial direction. 
The two independent Einstein equations then take the form (written in terms of geometrical radius $r(\rho)$)
\be
2rr''+r'^2-1 &=& 0 \,, \nonumber \\
\Omega(r'^2-1) + 2rr'\Omega' &=& 0 \, .
\lb{3}
\ee
\noindent {\bf Horizon in classical theory (aspect B):} {\it Suppose that the 2-sphere at $\rho=\rho_h$ is a minimal area surface, i.e. $r'=0$ at $\rho=\rho_h$. Then it follows from the 
second equation in (\ref{3}) that  the function $\Omega(\rho_h)=0$ and, hence, $\rho=\rho_h$ is a horizon.} Notice that $\Omega'(\rho_h)$ is not determined from (\ref{3}), it is a constant of integration. Fixing the temperature as the periodicity in the Euclidean time $t$, we can determine $\Omega'(\rho_h)$ by the condition of absence of a conical singularity 
in the metric (\ref{1}). Of course, globally, equations (\ref{3}) describe nothing else but the Schwarzschild solution.

\medskip

Below we shall examine the validity of the analogous aspects in the semiclassical gravity.

\section{Semiclassical gravitational equations}

The semiclassical gravitational action is composed by adding to the classical Einstein-Hilbert action $W_{EH}[G]$ a quantum effective action  $\Gamma[G]$ obtained by integrating out
the quantum matter fields. For simplicity we shall consider  conformal fields. In this case the difference between the effective actions for conformally related metrics, $\Gamma[e^{2\sigma}g]-\Gamma[g]$, is completely determined by the conformal anomaly \cite{anomaly}. Then, writing the metric in the form
$G_{\mu\nu}=e^{2\sigma}{g}_{\mu\nu}$ we find for the complete  gravitational action, $W_{grav}=W_{EH}[G]+\Gamma[G]$,
\be
W_{grav}&=&-\frac{1}{2\kappa}\int d^4x\sqrt{{G}}\,{R(G)}-\frac{a}{(4\pi)^2}\int d^4x\sqrt{g}\,\sigma\,C^2 +\frac{b}{(4\pi)^2}\int d^4x\sqrt{g}\,\sigma\,E \nonumber\\
&&-\frac{2b}{(4\pi)^2}\int d^4x\sqrt{g}\Big(2E^{\mu\nu}\nabla_\mu\sigma\nabla_\nu\sigma+2\,\Box\sigma\nabla_\mu\sigma\nabla^\mu\sigma + (\nabla_\mu\sigma\nabla^\mu\sigma)^2\Big)+\Gamma_0[{g}_{\mu\nu}]\,.\quad
\lb{4}
\ee
Here $\kappa = 8\pi G_N$ is the classical gravitational coupling, $E_{\mu\nu}$ is the Einstein tensor corresponding to the metric $g$, i.e. $E_{\mu\nu}=R_{\mu\nu}-\frac{1}{2}Rg_{\mu\nu}$ and $a$ and $b$ are the well known conformal charges, 
\begin{eqnarray}
a &=& \frac{n_0}{120}+\frac{n_{1/2}}{20}+\frac{n_1}{10} \,, \nonumber\\
b &=&\frac{n_0}{360}+\frac{11 n_{1/2}}{360}+\frac{31 n_1}{180}\,,
\lb{5}
\end{eqnarray}
where $n_s$ is number of fields of spin $s$.
We define
\be
C^2 &=& R_{\alpha\beta\mu\nu}R^{\alpha\beta\mu\nu}-2R_{\alpha\beta}R^{\alpha\beta} +\tfrac{1}{3}R^2 \,, \nonumber \\
E &=& R_{\alpha\beta\mu\nu}R^{\alpha\beta\mu\nu}-4R_{\alpha\beta}R^{\alpha\beta} +R^2 \,,
\lb{6}
\ee
where $E$ is the Euler density and $C$ is the Weyl tensor. 
We shall consider the gravitational action (\ref{4}) on the static metric (\ref{1}) so that $\Omega^2=e^{2\sigma}$ and $\Gamma_0[{g}_{\mu\nu}]$ is  the quantum effective action computed on the optical metric $g_{\mu\nu}$. It is interesting that the Euler density vanishes for the optical metric
(\ref{1}), $E(g)=0$. 
After integrating over $\theta$, $\phi$ and taking the periodicity in the Euclidean time to be $\beta$ we obtain the gravitational action for metric (\ref{1}) in the form
\begin{align}
&(4\pi\beta)^{-1}W_{grav} \\
&=\frac{1}{\kappa}\int dz\,\frac{e^{2\sigma}}{N^2}\Big(R'^2N+6RR'N\sigma'+3R^2N(\sigma'^2+\sigma'')+2RR''N -N^3-2RR'N'-3R^2N'\sigma'\Big) \nonumber \\
& \hspace{0.4cm} - \frac{4a}{3(4\pi)^2}\int dz\,\frac{\sigma}{R^2N^5}\left(N^3+RR''N- RR'N'-R'^2N\right)^2\nonumber \\
&\hspace{0.4cm} -\frac{4b}{(4\pi)^2}\int dz\left[\frac{1}{N}\left(\frac{R'^2}{N^2}-1\right)\,\sigma'^2 + \frac{R^2\sigma'^2}{N^3}\left(\sigma''+2\frac{R'}{R}\sigma'-\frac{N'}{N}\sigma'\right) + \frac{R^2\sigma'^4}{2N^3} \right]+(4\pi\beta)^{-1}\Gamma_0[g_{\mu\nu}] \,,\nonumber
\end{align}
where $\displaystyle \sigma'\equiv\frac{d\sigma}{dz}$.
The variations with respect to $\sigma(z)$ and $N(z)$ give us the equations of motion (the third equation obtained by variation with respect to $R(z)$, is supposed to follow from these two
by the Bianchi identities):
\begin{align}
0= &\frac{2e^{2\sigma}}{\kappa}\left[\frac{2 R R''}{N}+\frac{6 R R' \sigma'}{N}-\frac{2 R R' N'}{N^2}+\frac{R'^2}{N}+\frac{3R^2 \sigma''}{N}-\frac{3 R^2 \sigma'N'}{N^2}+\frac{3 R^2 \sigma'^2}{N}-N\right]\lb{8}\\
&+\frac{a}{6\pi^2}\left[-\frac{R''}{RN}-\frac{R''^2}{2 N^3}-\frac{R'^3 N'}{R N^4}-\frac{R'^2 N'^2}{2 N^5}+\frac{R' N'}{R N^2}-\frac{R'^4}{2 R^2 N^3}+\frac{R'^2}{R^2 N}+\frac{R'R'' N'}{N^4}+\frac{R'^2 R''}{R N^3}-\frac{N}{2 R^2}\right]\nonumber\\
&+\frac{b}{\pi^2 N^4}\Bigg[RN R''\sigma'^2+\frac{1}{2} NR'^2 \sigma''-3 R R' \sigma'^2 N'-\frac{3}{2} R'^2 \sigma'N'+R NR' \sigma'^3+N R'^2 \sigma'^2\nonumber\\
&+2 R N R' \sigma' \sigma''+N R' R''\sigma'-\frac{3}{2} R^2 \sigma'^3 N'+\frac{3}{2} R^2 N \sigma'^2 \sigma''-\frac{1}{2} N^3 \sigma''+\frac{1}{2} N^2 \sigma' N'\Bigg]\nonumber
\end{align}
\begin{align}
0= &\frac{e^{2\sigma}}{\kappa N^2}\left[\left(R'+R\sigma'\right) \left(R'+3 R \sigma'\right)-N^2\right]\lb{9}\\
&+\frac{b\sigma'^2}{8\pi^2N^4}\left[-2 N^2+8 R R' \sigma'+6 R'^2+3 R^2 \sigma'^2\right]\nonumber\\
&+\frac{a}{12\pi^2R^2}\Bigg[\frac{R^2 \sigma R''^2}{N^4}+\frac{2 R^2 R'^2 \sigma' N'}{N^5}+\frac{2 RR'^3 \sigma '}{N^4}-\frac{2 R R' \sigma '}{N^2}+\frac{2 R^2 \sigma R'^2 N''}{N^5}\nonumber\\
&-\frac{5 R^2 \sigma R'^2 N'^2}{N^6}+\frac{\sigma R'^4}{N^4}-\frac{2 R^2 R''' \sigma R'}{N^4}-\frac{2 R^2 R' R'' \sigma '}{N^4}+\frac{4 R^2 \sigma R' R''N'}{N^5}-\sigma\Bigg]+\frac{1}{4\pi \beta}\delta_N\Gamma_0\nonumber
\end{align}
These are the equations which we shall further analyze.

\section{Effective action $\Gamma_0$ on optical space $S_1^\beta\times M_3$}

Before we do this analysis, we have to discuss $\Gamma_0$ that appears in (\ref{9}). $\Gamma_0[g]$ is the quantum effective action computed on the
optical metric $g_{\mu\nu}$,
\be
ds^2(g)=dt^2+ds^2(\gamma)\, , \ \ ds^2(\gamma)=N^2(z) dz^2+R^2(z)(d\theta^2+\sin^2\theta d\phi^2)\, ,
\lb{10}
\ee
where the $t$-coordinate is compact with periodicity $\beta=1/T$. Thus, the optical spacetime is a product $S_1^\beta\times M_3$. The effective action on this spacetime
can be decomposed in a series in terms of curvature of 3-space $M_3$ \cite{Gusev:1998rp}. Restricting to the leading terms in this expansion and combining results of
\cite{Gusev:1998rp} with those of \cite{S1H3} we obtain
\be
\Gamma_0[S_1^\beta\times M_3]=-\frac{\pi^2}{90\beta^3}\left(n_0+\frac{7}{2}n_{1/2}+2n_1\right)\int_{M_3}1+\frac{1}{144\beta}\left(n_{1/2}+4n_1\right)\int_{M_3}{\cal R}_M\, , 
\lb{10-0}
\ee
where 
\[{\cal R}_M=-\frac{2}{R^2N^3}\left(2RNR''-2RR'N'-N^3+NR'^2\right)
\]
is the Ricci scalar of $M_3$.
Notice that in the near horizon limit ($z\rightarrow \infty$) (\ref{2}), in the gauge $N(z)=1$,
the 3-space $M_3$ approaches asymptotically the hyperbolic Euclidean space $H_3$, as was noticed earlier in \cite{Sachs:2001qb}. The radius of the hyperbolic space 
$H_3$, $a=\beta/(2\pi)$, is related to the inverse temperature, or equivalently, to the periodicity $\beta$ of the circle $S_1$. In the case, where $M_3=H_3$,
(\ref{10-0}) is the complete result for the effective action. The higher order corrections to (\ref{10-0}) thus vanish if $M_3$ has constant curvature.
After a number of integrations by parts in the second term in (\ref{10-0}) we arrive at
\be
\Gamma_0[S_1\times H_3]=-\frac{2\pi^3}{ 45}\frac{c_H}{\beta^3}\int dz N(z)R^2(z)+\frac{\pi}{18}\frac{\lambda_H}{\beta} \int dz(N(z)+R'^2N^{-1})\, , 
\lb{11}
\ee
where we introduced $c_H=n_0+\frac{7}{2}n_{1/2}+2n_1$ and $\lambda_H=n_{1/2}+4n_1$. So that one finds\footnote{It is worth noting that a priori the thermal action $\Gamma_0$
is not supposed to be determined by only the conformal properties of the theory. Therefore, it is quite interesting that there exists a certain relation between
charges $a$ and $b$ and the parameters appearing in $\Gamma_0$: $180b+60a+90c-c_H-5\lambda_H=0$, where $c=-n_1/6$. In certain regularizations
combination $(2/3a+c)$ appears in front of $\Box R$  in conformal anomaly. This term originates from a local term in the effective action, depends on the regularization scheme and the gauge fixing condition and does not affect 
our analysis below in the paper. We thank the anonymous referee for pointing out the above relation to us.} 
\be
(4\pi\beta)^{-1}\delta_N\Gamma_0=-\frac{\pi^2}{90}\frac{c_H}{\beta^4}R^2(z)-\frac{1}{72}\frac{\lambda_H}{\beta^2}(R'^2N^{-2}-1)
\lb{11-1}
\ee
that should be used in (\ref{9}).

\section{No static black hole solutions of semiclassical equations}
We are now in a position to analyze the semiclassical equations (\ref{8})-(\ref{9}). First we analyze  the aspect A.  Here we use the gauge $N(z)=1$. We are looking at a solution which describes a static spherically symmetric geometry
(\ref{1}) with a horizon characterized by a finite radius $r_h$ and a finite Hawking temperature $T=1/\beta$. So that the  functions $\sigma(z)$ and $R(z)$ behave to leading order ($z\rightarrow\infty$) as
\be
\sigma(z)=-\frac{2\pi z}{\beta} +\dots \, , \quad R(z)=r_h e^{2\pi z/\beta}+\dots
\lb{12}
\ee
and one has asymptotically that $R'\simeq(2\pi/\beta)R$.
Then, to leading order, equations (\ref{8}) and (\ref{9}) respectively give
\be
\delta_\sigma W_{grav}: \quad 0&=& {\cal O}\left(e^{-4\pi z/\beta}\right), \lb{13}\\
\delta_N W_{grav}: \quad 0&=& \left(360b-2c_H-10\lambda_H\right)\frac{\pi^2}{180\beta^4}R^2(z)=-\left(n_0+6n_{1/2}-18n_1\right) \frac{\pi^2}{180\beta^4}R^2(z)\, .\nonumber
\ee
First equation in (\ref{13}) is automatically satisfied to leading order. Although the divergent term in the second equation may vanish for a particular set of fields,
it is not vanishing in general, for arbitrary $n_0$, $n_{1/2}$ and $n_1$.
Obviously, since the leading divergent term is not cancelled, equation (\ref{13}) can not be satisfied and, hence, a static solution with a finite temperature horizon can not exist in the semiclassical gravity theory.
This is our first important observation. 

\section{Minimal sphere is a wormhole throat}
Now, as the classical aspect A has, in general,  failed in the semiclassical theory  we want to check the second aspect (B)  of the classical black hole horizons. 
Namely, we want to check whether  a minimal sphere  is necessarily a black hole,  now for
the semiclassical gravity described by equations (\ref{8}) and (\ref{9}). It is natural in this case to choose in these  general equations the gauge, where $N(z)=1/\Omega(z)$, so that coordinate $\rho=z$ would correspond to the geodesic distance in the radial direction. 
One possibility is to look at the solutions with $r'=0$, $\Omega=0$, $\Omega'=2\pi/\beta$. Those solutions do not in general exist by same reasons as above (uncancelled $1/\Omega^2$ divergent terms). Under same conditions as non-vanishing of (\ref{13}), this excludes the possibility that the minimal surface is a horizon. 
Therefore, we generalize the minimality condition and look for solutions in which both $r(\rho)$ and $\Omega(\rho)$ (or, equivalently, $G_{tt}$)  have minimum at $\rho=\rho_h$.
Thus, we  impose two conditions at the turning point $\rho=\rho_h$, $r'(\rho_h)=0$ and $\Omega'(\rho_h)=0$. 
In the present case it is natural to analyze the equations in terms of the geometrical radius of the sphere $r(\rho)=R(\rho)\Omega(\rho)$ and the function $\Omega(\rho)=e^{\sigma(\rho)}$. Then, equations (\ref{8}) and (\ref{9}) at the turning point take the form
\be
\frac{2\om}{\kappa}\left(1-2rr''-r^2 \frac{\om''}{\om}\right)+
\frac{\ba} {r^2 \Omega}\left(\Omega+\om r r''-r^2 \om''\right)^2+\bb\om''=0\,,
\lb{14}
\ee
\be
-\frac{\om^2}{\kappa}-\frac{\ba}{r^2} \ln \om^{-1}\left[(\om r r''-r^2\om'')^2-\om^2\right]-\frac{\gamma r^2}{\beta^4\om^2}+\frac{{\lambda}}{\beta^2}=0\, ,
\lb{15}
\ee
where $\ba=a/12\pi^2$, $\bb=b/2\pi^2$, $\gamma=c_H\pi^2/90$ and ${\lambda}=\lambda_H/72$, and we note $\Omega\equiv\Omega(\rho=\rho_h)$, $r\equiv r(\rho=\rho_h)$.
Notice, that in the classical limit $r$ is the geometrical radius of the  black hole.

 The turning point is assumed to be a minimum both for functions $r(\rho)$ and $\Omega(\rho)$ so that
their second derivatives $r''>0$ and $\Omega''>0$ at $\rho=\rho_h$. Additionally, we require that $0<\Omega<1$ since it is expected to be a small modification
of the classical value $\Omega (\rho_h)=0$. 
Let us introduce a new variable $y$ such that the equation (\ref{15}) can be rewritten as follows
\be
&&\left(\om r r''-r^2\om''\right)^2=y^2\om^2 \, , \nonumber \\
&&y^2=1-\frac{r^2}{\kappa\bar{a}\ln\om^{-1}}\left(\frac{\gamma\kappa r^2}{\beta^4\om^4}-\frac{{\lambda}\kappa}{\beta^2\om^2}+1\right)\, . \ \ 
\lb{16}
\ee
The positivity condition $y^2\geq 0$ imposes important constraints on possible values of $\Omega$ and $\beta$.

\medskip
\noindent
{\bf First, we consider the case when $\lambda=0$. }Then since $a>0$, $b>0$, $\gamma>0$, and $0<\om<1$ one has that $y^2<1$. 
On the other hand, the positivity condition $y^2>0$ can be rewritten in the form of an inequality
\be
\om^4\ln\frac{\om_0}{\om}>\frac{\gamma r^4}{\ba\beta^4}\, ,
\lb{17}
\ee
where we introduced $\om_0=e^{-\frac{r^2}{\ba \kappa}}$. This condition is very informative.  First of all it says that provided the temperature is finite, $1/\beta\neq 0$,
the norm of the Killing vector $\om^2$ at the turning point  does not vanish. 
Then, eq. (\ref{17}) implies that
\be
\Omega<\om_0=e^{-\frac{r^2}{\ba \kappa}}\, .
\lb{18}
\ee
This relation indicates that the value of $\Omega$ at the turning point is bounded by the exponential of minus the classical Bekenstein-Hawking
entropy $S_{BH}=8\pi^2 r^2/\kappa$. 

On the other hand, (\ref{17}) can be viewed as a bound on the possible temperature,
\be
T^4=\frac{1}{\beta^4}<\frac{\ba}{\gamma r^4}\om^4\ln\frac{\om_0}{\om}<\frac{1}{4}\frac{\ba}{\gamma r^4}\om_0^4\, ,
\lb{19}
\ee
where in the last inequality we used that  $\Omega^4\ln \om_0/\om\leq \om_0^4/(4e)\leq \om_0^4/4$. This relation indicates that
the temperature of the semiclassical geometry that replaces the classical black hole is much less than the Hawking temperature. 

\medskip
\noindent
{\bf Now we consider the case of non-vanishing $\lambda>0$.} The positivity condition $y^2\geq 0$ in this case implies the inequality
\be
\om^4\ln \frac{\om_0}{\om}\geq \frac{\gamma r^4}{\ba \beta^4}\left(1-\frac{\lambda}{\gamma}\frac{\beta^2}{r^2}\om^2\right)\,.
\lb{25}
\ee
There are different sub-cases that one should consider. \\
\noindent {\bf Case A.} Suppose that the right hand side of (\ref{25}) is positive, i.e. $\om^2<\frac{\gamma}{\lambda}\frac{r^2}{\beta^2}$.
Then it follows that one again has the bound (\ref{18}), $\om<\om_0$. Using that $\om^4\ln\frac{\om_0}{\om}\leq \frac{1}{4}\om_0^4$ one arrives at the inequality
\be
\beta^4 \frac{1}{4}\om_0^4+\frac{\lambda}{\ba}r^2\om_0^2\beta^2-\frac{\gamma}{\ba}r^4\geq 0
\lb{26}
\ee
that should be considered as a restriction on the possible values of $\beta$. Solving the quadratic equation one finds two roots,
\[\beta_{1,2}^2=\frac{2\lambda}{\ba}\frac{r^2}{\om_0^2}\left(\pm\sqrt{1+\frac{\gamma\ba}{\lambda^2}}-1\right),\qquad \beta_1^2<\beta_2^2.
\] 
We see that $\beta_1^2<0$ while
$\beta_2^2>0$. Therefore, (\ref{26}) implies that $\beta^2>\beta_2^2$, or equivalently,
\be
T^2=\frac{1}{\beta^2}<\frac{\ba \om_0^2}{2r^2}\frac{1}{\sqrt{\lambda^2+\gamma\ba}-\lambda}\, .
\lb{27}
\ee
When $\lambda=0$ this bound becomes (\ref{19}).

\noindent {\bf Case B.} The other possibility is when the right hand side of (\ref{25}) is negative, i.e. $\om^2>\frac{\gamma}{\lambda}\frac{r^2}{\beta^2}$.
In this case the ratio $\om_0/\om$ can either be larger or smaller than $1$.
If we assume that $\om\leq \om_0$, then the analysis above is still valid and one has an upper bound (\ref{27}) on the temperature $T$. However, this time we have 
$\frac{1}{\beta^2}<\frac{\lambda}{\gamma}\frac{\om^2}{r^2}$ and since $\om<\om_0$ we obtain a stronger bound than (\ref{27})\footnote{That the bound (\ref{28}) is stronger than (\ref{27}) follows from the fact that the function $\frac{x}{2(\sqrt{1+x}-1)}>1$, $x=\gamma\ba/\lambda^2$.}
\be
T^2=\frac{1}{\beta^2}<\frac{\lambda}{\gamma}\frac{\om^2_0}{r^2}\, .
\lb{28}
\ee

On the other hand, if $\om>\om_0$ then (\ref{25}) can be rewritten in the form
\be
\beta^4\om^4\ln\frac{\om}{\om_0}-\frac{\lambda}{\ba}r^2\om^2\beta^2+\frac{\gamma}{\ba}r^4 <0
\lb{29}
\ee
that again should be considered as a restriction on possible values of $\beta$. Solving the quadratic equation one finds
for the roots 
\[
\beta^2_{3,4}=\frac{\lambda r^2}{2\ba\om^2\ln\frac{\om}{\om_0}}\left(1\pm\sqrt{1-\frac{4\gamma\ba}{\lambda^2}\ln \frac{\om}{\om_0}}\right).
\]
The necessary condition for the inequality (\ref{29}) to have a non-trivial domain of validity is that $\frac{4\gamma\ba}{\lambda^2}\ln\frac{\om}{\om_0}\leq1$ or, equivalently,
\be
\om_0<\om<\om_0 e^{\frac{\lambda^2}{4\gamma\ba}}\, .
\lb{30}
\ee
We see that in this case there is not only a lower but also an upper bound on the possible value of $\Omega$ at the throat, which are exponentially small functions of the classical BH entropy.
One also finds that both the roots are real and positive $\beta_3^2<\beta_4^2$, $\beta^2_{3,4}>0$. The domain where the condition (\ref{29}) is satisfied is then
$\beta^2_3<\beta^2<\beta^2_4$, which imposes both upper and lower limits on the temperature. A somewhat simpler upper bound on the temperature can be derived by using the condition $\om^2>\frac{\gamma}{\lambda}\frac{r^2}{\beta^2}$. Indeed one finds that
\be
T^2=\frac{1}{\beta^2}<\frac{\lambda}{\gamma}\, e^{\frac{\lambda^2}{2\gamma\ba}}\, \frac{\om_0^2}{r^2}\, .
\lb{31}
\ee
\medskip
\noindent {\bf The minimality conditions:} Equations (\ref{14}) and (\ref{15}) can be used to express $r''$ and $\om''$ in terms of variable $y$,
\be
\om''=\frac{\om}{r^2\left(3-\frac{\bb\kappa}{2r^2}\right)}\left(1-2y+\frac{\ba\kappa}{2r^2}(1+y)^2\right),
\lb{20}
\ee
\be
rr''=\frac{1}{\left(3-\frac{\bb\kappa}{2r^2}\right)}\left[\left(1+\frac{\ba\kappa}{2r^2}\right)+\left(1-\frac{\left(-\ba+\frac{\bb}{2}\right)\kappa}{r^2}\right)y+\frac{\ba\kappa}{2r^2}y^2\right] .
\lb{21}
\ee
We look at large black holes, $\kappa/ r^2 \ll 1$, so that $\left(3-\frac{\bb\kappa}{2r^2}\right)>0$.
The condition $\Omega''>0$ then imposes a condition on the possible values of $y$:
\be
y<y_1\ \ {\rm or} \ \ y>y_2\, , \ \ y_1=\frac{1}{2}+\frac{9}{16}\frac{\ba\kappa}{r^2}\, , \ \ y_2=4\frac{r^2}{\ba \kappa}\, ,
\lb{22}
\ee
where we skip the subleading in $\kappa/r^2$ terms. 
On the other hand, the condition $r''>0$ gives us the conditions
\be
y<y_3 \ \ {\rm or} \ \ y>y_4\, , \ \ y_3=-\left(1+\frac{\bb\kappa}{2r^2}\right)\, , \ \ y_4=-\frac{2r^2}{\ba\kappa}\, .
\lb{23}
\ee
There are three  intersections of regions (\ref{22}) and (\ref{23}) 
\be
(I): \ \ y<y_3\, , \ \  (II): \ \   y_3<y<y_1\, , \ \ (III): \ \ y>y_2\, .
\lb{24}
\ee
When $\lambda=0$ and in the case A when $\lambda>0$ one has that $y^2<1$ so that the regions I and III are not admissible while  region II reduces to a somewhat smaller, $-1<y<y_1$. On the other hand, in the case B when $\lambda>0$ one has that $y^2>1$ and the only admissible regions are I and III and a part of region II, $y_3<y<-1$.

\medskip

The relations (\ref{18}) and (\ref{30}) imply that at the turning point the norm $\xi^2=\Omega^2$ of the time-like Killing vector $\xi=\partial_t$ is an exponentially small function of the 
Bekenstein-Hawking entropy of the classical black hole. The minimal 2-sphere is then not a horizon as it was in the classical case but a wormhole.
This is our second important observation. It answers the question what replaces the classical horizon in the semiclassical theory.
Additionally, we have found restrictions (\ref{19}), (\ref{27}), (\ref{28}), (\ref{31}) on the possible values of the temperature. In all these cases, the temperature happens to be
exponentially small and, thus, differs radically from the Hawking temperature.

\section{Remarks}\label{sec:remarks}

\begin{enumerate}

\item The analysis of the present paper is done in the approximation when the gravity is not quantized. For quantum conformal matter, the effective action we consider is a one-loop result. If the conformal fields are non-interacting, this is in fact the \emph{entire} answer for the effective action. If the fields are interacting, then there are higher loop corrections.
One effect of interactions is that the conformal charges $a$ and $b$, may RG flow, i.e. depend on the scale. It would be interesting to explore the consequences of this flow in the
situation at hand. Another way to generalize our analysis is to consider massive fields. 

\item   The other essential approximation we used in this paper is the approximation of the effective action $\Gamma_0$ on the optical metric
by expression (\ref{10-0}) that ignores the possible higher curvature terms. One can estimate the order of the possible correction terms to (\ref{10-0}).
The optical geometry (\ref{10}) is characterized by two dimensionfull quantities: $\beta$ and $R(z)$. The possible correction terms will then go as a ratio
$\beta/R(z)\sim \Omega(z)$. This ratio vanishes at the true horizon, which explains why (\ref{10-0}) is the complete result for the near-horizon geometry $S_1\times H_3$.
On the other hand, this ratio is exponentially small at the wormhole throat. The corrections thus are expected not to affect the horizon analysis leading to eq.(\ref{13})
and to be small in the analysis of the wormhole.  

\item Modulo the situation discussed in the above points, our consideration is essentially \emph{non-perturbative} and does not rely on any perturbative analysis of the semiclassical equations. This is an important difference between our approach and other previous attempts
to understand what happens to classical black holes in a quantum theory.

\item\label{stateremark} Our results differ from the known considerations of black holes in local theories of gravity. No local curvature invariant, considered on a static metric of the type (\ref{1}), in the gravitational action will produce a divergent term like (\ref{13}).  It is the fundamental non-locality of the quantum effective action, even though it is manifested via a local form of the conformal anomaly, that lies at the root of (\ref{13}). 

In order to illustrate this point we consider a local $R^2$ term which always can be added to the effective action.
After conformal transformation $G_{\mu\nu}=e^{2\sigma}g_{\mu\nu}$ it produces
\be
\int \sqrt{G}R^2(G)=12\int \sqrt{g}\left[\frac{1}{12}R^2-R\Box\sigma -R (\nabla\sigma)^2+3(\Box\sigma)^2+6\Box\sigma (\nabla\sigma)^2+3(\nabla\sigma)^4\right] .
\lb{AB}
\ee
In fact, as is well known, this term produces an extra contribution to the conformal anomaly, proportional to $\Box R$ (as one can see from (\ref{AB}) after 
few integrations by parts are made).  Now, the term (\ref{AB}) should be considered for the class of metrics (\ref{1}) (note that $\Omega=e^\sigma$) so that it becomes a functional of 
3 functions: $\sigma(z), \, N(z), \, R(z)$. Variations of (\ref{AB}) w.r.t. $\sigma(z)$ and $N(z)$ represent the modifications of the semiclassical equations discussed
in this paper. We have however checked that the modifications originated from (\ref{AB}) and considered on the class of functions satisfying the asymptotic 
conditions (\ref{12}) do not produce any divergent terms of the type that appear in (\ref{13}). In fact these modifications asymptotically vanish as $O(e^{-4\pi z/\beta})$.
This is so even though each individual term in (\ref{AB}) may have a divergence proportional to $R^2(z)\sim e^{4\pi z/\beta}$, simply these divergences
are mutually cancelled in the entire expression (\ref{AB}). We expect this cancellation to happen for any local invariant constructed from metric and its derivatives. In fact, as is well known \cite{local_refs}, these local terms associated with $\Box R$ contributions are non-universal and regularization scheme dependent. Moreover, the coefficients with which the Maxwell field contributes to the trace anomaly, also depends on the choice of gauge. Thus, even though one can not exclude a suitably chosen combination of local terms which might cancel out our divergence structures that we discussed in (\ref{13}), we should exclude such terms for all physical purposes.

\item It is well known \cite{BD} that in the Boulware vacuum, which corresponds to a static asymptotically flat spacetime,  the expectation value of stress-energy tensor 
is divergent on the horizon.  Therefore, it is rather natural to expect that in this state, provided the back-reaction is consistently taken into account, the horizon would not exist
and would be replaced by something else. Although we did not specify explicitly the quantum state, the choice appears to be made as soon as we use the `hyperbolic' boundary conditions
while computing $\Gamma_0$.  Our analysis provides a self-consistent treatment of the back-reaction problem and one of our principal results  rigorously and quantitatively shows that in the complete theory, the horizon is replaced by  the wormhole throat. To the best of our knowledge the divergence in (\ref{13}) for fields of various spin was not reported before in the literature.

\item In the classical limit, the charges $a$ and $b$ are taken to zero. In this limit, $\Omega(\rho_h)$ vanishes and the wormhole solutions, we have found, become black holes.
The temperature, however, does not go analytically to the ``classical'' Hawking temperature because the latter is fixed by regularity in the near-horizon geometry and since the topology of the wormhole and black hole geometries are different, the regularity requirement does not arise for the wormholes.

\item As our analysis shows, for a generic set of fields, the presence of the divergent term in equation (\ref{13}) eliminates the solutions with black hole horizons.
However, for certain sets of fields this divergent term can be canceled, e.g. if there are $n_0=6$ scalar fields, $n_{1/2}=2$ Dirac fermions and $n_1=1$ vector field. Quite surprisingly, this is precisely the multiplet of ${\cal N}=4$ super-Yang-Mills theory in $4$ dimensions. It may mean that in this case a black hole solution exists, demonstration of which may require the analysis of the next to leading order terms in the gravitational equations.  A signal of a similar phenomena has been observed earlier within the AdS/CFT correspondence, \cite{AdS/CFT}.  As is transparent from our analysis the possible cancellation should be attributed to the particular field content rather than to the strength of interactions.   One should also note that the cancelation condition of the divergence (\ref{13}) is not in general equivalent to the condition $a=b$ for the conformal charges that is typical for the holographic theories in $4$ dimensions. 

\item In GR with matter satisfying a positive energy condition (PEC), the metric component $g_{tt}=\Omega^2$ is a monotonic function of the radial coordinate, see for instance
\cite{BGS}, so that there is only one local minimum. In semiclassical gravity, the PEC effectively does not hold and one can not exclude the existence of multiple minima.
Our analysis shows that the value of $\om$ at each minimum is less than $\om_0$. We are interested in a wormhole solution which is asymptotically flat, i.e. $\Omega=1$ at infinity.
In this solution, as soon as the barrier set by $\om_0$ is passed, no new minimum will appear and the function $\om$ will grow monotonically to 1 in the radial direction.

\item An interesting question is whether the wormhole solution which corresponds to a given classical black hole metric with the horizon radius $r$ is unique.
The wormhole solutions are parametrized by the variable $y$ which may take its value in one of the regions defined in (\ref{24}). Thus, for a given value of $r$
and upon condition of approaching Minkowski spacetime at infinity, there may be a family of the corresponding wormholes. A more precise specification of this family requires a more detailed analysis of the global structure of the wormhole solutions.

\item On the observational side, the question whether the numerous observable astrophysical candidates for black holes are actual black holes and not something else which only look like
black holes, becomes more and more of a practical nature. In this respect we repeat here the point made in \cite{Damour:2007ap}. The characteristic observation time needed to distinguish the wormhole mimicker from an actual black hole of mass $M$ is of the order $t\sim G_N M\ln (1/\om_0)$, where $\om_0^2$ is the value of $G_{tt}$ in the throat.
For $\om_0$ exponentially small function of the BH entropy this time $t\sim G_N^2 M^3$ is of order of the evaporation time of a black hole. Clearly, this time is too long to 
be of any relevance to any actual observations. Unless there is a way to compress this time to something much shorter, no experiment or observation would be able to 
conclusively prove the (non)-existence of a black hole horizon. For now it seems that the only decisive manifestation of the actual black holes would be a direct detection of the Hawking radiation with the predicted Hawking temperature. For astrophysical objects this is of course not a feasible option. 

\item Since we predict that the possible temperature of the wormholes that mimic the black holes is much smaller than the Hawking temperature, they should live much longer
than the actual black holes. Thus, any small ``black holes'' that were formed at the early stages of the Universe should not completely evaporate by now, but survive till the present epoch. Their observation would indirectly confirm our predictions.

\item  In this note we are mostly interested in large black holes, the area of which is much larger than the Planck area $r^2/\kappa \gg 1$. However, our basic formulas
are valid for any radius $r$ and, hence, it would be interesting to analyze the implications of our approach to the Planckian size black holes, when $r^2\sim \kappa^2$.
This work is in progress.

\item Obviously, the existence of wormholes instead of black holes suggests a new way to resolve the old problem (puzzle) of the information loss \cite{infoloss}, \cite{Solodukhin:2004rv}, \cite{Germani:2015tda}. This, however, goes beyond the scope of the present paper.

\end{enumerate}

\section*{Acknowledgements} S.S. would like to thank A. Barvinsky, A. Zelnikov and M. Parikh for useful discussions on various stages of this project.
The preliminary results were reported by S.S. at the workshop ``Cosmology and High Energy Physics III", Montpellier, 15 December 2017.
He thanks Sergei Alexandrov and David Polarski for their kind hospitality at the University of Montpellier. D.S. is supported through the NCCR SwissMAP (The Mathematics of Physics) of the Swiss Science Foundation. He thanks Matthias Blau for some helpful discussions.

\end{document}